# Evaluating and predicting the Efficiency Index for Stereotactic Radiosurgery Plans using RapidMiner GO(JAVA) Based Artificial Intelligence Algorithms.

**Hossam Donya[1,2*], Sheikh Othman,[1] Alexis Dimitriadis[3]**

[1]Department of Physics, Faculty of Science, King Abdulaziz University, Jeddah 21589, Saudi Arabia

[2]Department of Physics, Faculty of Science, Menoufia University, Shibin El-Koom, Egypt

[3]Queen Square Radiosurgery Centre, National Hospital for Neurology and Neurosurgery, London, United Kingdom

**Correspondence: Hossam Donya:** *Department of Physics, Faculty of Science, King Abdulaziz University, Jeddah 21589, Saudi Arabia.* hdunia@kau.edu.sa

## *Abstract*

***Objective***: We analyze a new method to evaluates the prediction of Efficiency index ($n_{50\%}$) by DVH parameter for stereotactic SRS treatment plans using Supervised Machine learning and evaluate the performance of predictive model algorithms of RapidMiner GO (9.8 version) in the parameter prediction.

***Methods:*** *Dose volume histogram (DVH) based Efficiency index ($n_{50\%}$) was calculated for 100 clinical SRS plans generated by Leksell Gamma plan, and the results were compared to predicted values produced by machine learning toolbox of RapidMiner Go, algorithms are namely, Generalized linear model (GLR), Decision Tree Model, Support Vector Machine (SVM), Gradient Boosted Trees (GBT), Random Forest (RF) and Deep learning Model (DL). Root mean square error (RMSE), Average absolute error, Absolute relative error, squared correlation and model building time were determined to evaluate the performance of each algorithm.*

***Results***: *The GLR algorithm model had square correlation of 0.974 with the smallest RMSE of 0.01, relatively high prediction speed, and fast model building time with 2.812 s, according to the results. The RMSE values for all models were between 0.01–0.021, all algorithms performed well. The RMSE of the Gradient Boosted Tree, Random Forest, and Decision Tree regression algorithms was found to be greater than 0.01, suggesting that they are not appropriate for predicting EI in this analysis.*

***Conclusions:*** *RapidMiner GO machine learning models can be used to predict DVH parameters like EI in SRS treatment planning QA. To effectively evaluate the parameter, it is necessary to choose a suitable machine learning algorithm.*

## *Introduction*

*Stereotactic radiosurgery (SRS) used a high dose of radiation with less fractionation for treating intracranial legions. The target is usually very small that it is often surrounded by healthy tissues of brain, which is considered as the primary organ at risk (OAR). To prevent normal tissue toxicity and complications, it's important to achieve a sharp dose falloff*

*from the target, irradiating less to the normal tissue in these procedures. High levels of adherence and steep dose gradients from the target's periphery to surrounding tissue are hallmarks of stereotactic radiosurgery (SRS). From data available it is evident that there are threshold doses at which the risk of symptomatic radio-necrosis rises in proportion to the amount normal tissue volume irradiated. As a result, SRS treatment plans should strive to reduce dose to surrounding tissue while increasing dose to the target amount. A number of metrics for assessing the quality of radio-surgical plans have been suggested, but each has flaws. So, a newly proposed novel metric Efficiency Index ($n_{50\%}$)[1, 2] is used in this study which is calculated the integral doses of required volumes.*

*The aim of this study is to evaluate the machine learning algorithms of RapidMiner GO, whether it can efficiently predict DVH parameters such as EI in SRS for treatment planning Quality Assurance.*

## Methods:

*Efficiency Index ($n_{50\%}$), The novel index is calculated as the ratio of the integral dose of TV to the integral dose of PIV50%:*

$$(n50\%) = \frac{Useful\ Energy}{total\ Energy} = \frac{Integral\ Dose\ TV}{Integral\ Dose\ PIV\ 50\%} = \frac{\int_{Dmin}^{Dmax} TV\ \delta dose}{\int_{PIV50\%}^{Dmax} V\ \delta dose} \ldots\ldots\ldots\ldots [1]$$

*Where, $D_{min}$ and $D_{max}$, is the minimum and maximum dose to the target volume (TV), and PIV50% is the volume occupied by 50% of the prescription dose (PD). EI($n_{50\%}$) for a treatment plan with a single target can be calculated using the integral dose to TV and of PIV50%, it can be achieved by using equation 4.[1]*

$$Integral\ Dose\ TV = Mean\ Dose\ TV \times Volume\ TV \ldots\ldots\ldots\ [2]$$

*This parameter is manually calculated from DVH exported from Leksell Gamma Plan (LGP). LGP is a powerful treatment planning software which is integrated with the Leksell gamma knife (LGK) and tools available performs Flawless workflow for complicated treatment plans.*

*In addition, EI($n_{50\%}$) is predicted considering the indices as well such as Paddick conformity index (PCI)(eqn 1) and Gradient index (GI)(eqn 2) and using parameters like treatment prescription isodose, coverage, selectivity, target volume and prescribed dose were used.[1]*

$$PCI = \frac{TV_{PIV}^2}{TV * PIV} \ldots\ldots\ldots [3]$$

*PIV (prescription isodose volume), TV (target volume). $TV_{PIV}$ (volume of the target covered by the prescription isodose).*

$$GI = \frac{PIV_{50\%}}{PIV_{100\%}} \ldots\ldots\ldots [4]$$

*Where, PIV50% and PIV100%, is the volume of 50% and 100% of the prescription isodose.[1]*

## Workflow of Machine Learning:

*RapidMiner Go (9.8) is an application based on JAVA Programming, it is available for educational analysis and research purpose. Provided with statistical tool and Machine Learning toolbox with several predictive algorithms. It requires minimum of 100 rows as a dataset to run any predictive analysis and perform the ''Regression'' modeling. It gives automated results and also allows you to make changes and creat our own model (figure 2).*

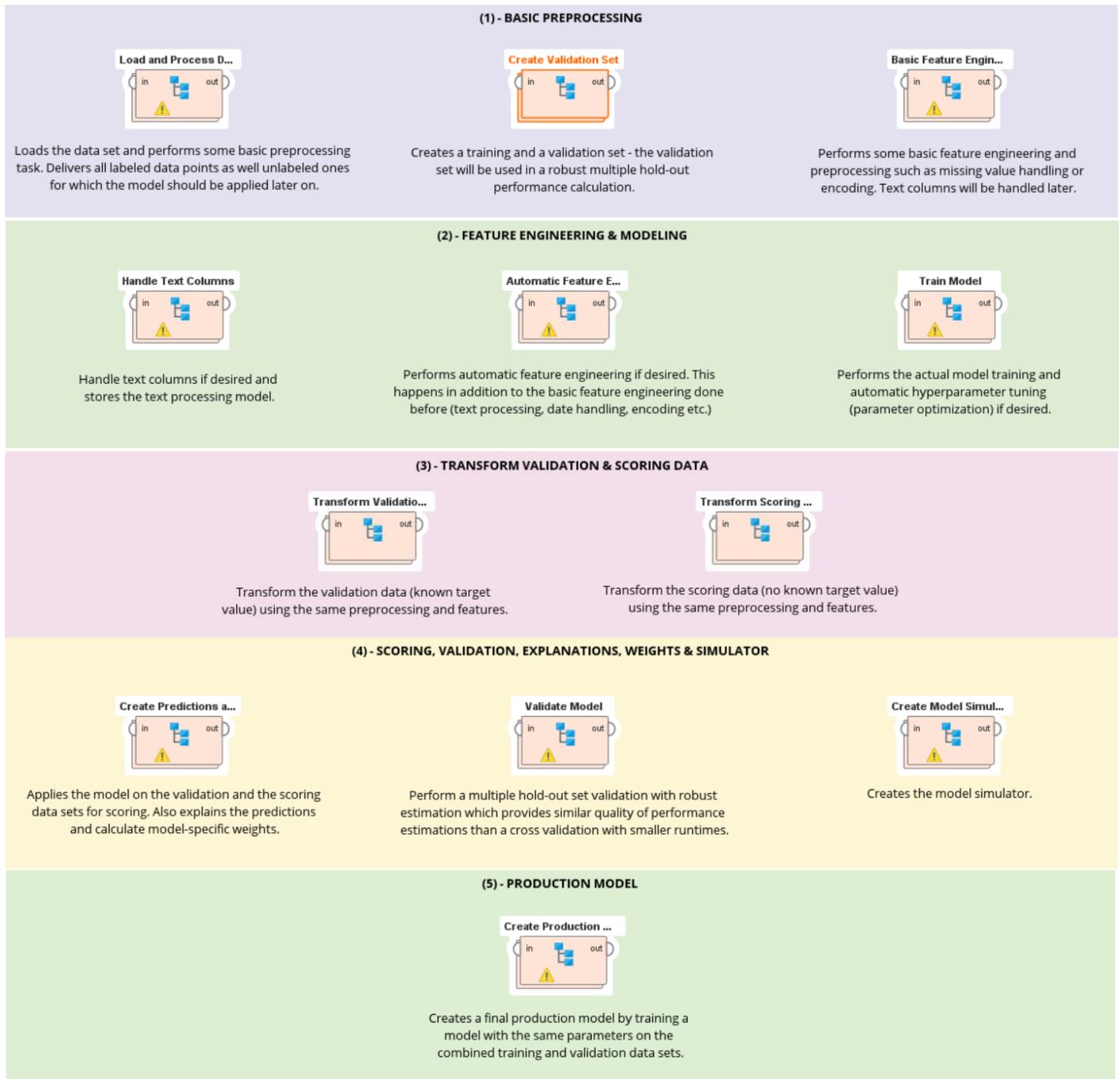

*Figure 1: Flow chart Model selection, training and Validation*

## Data collection and preprocessing

The value of EI($n_{50\%}$) and other supporting indices was calculated for 100 clinical intracranial SRS treatments that were reported in our previous reference [2] planned on the Leksell Gamma-Plan in order to examine the value of EI (n50%)[2]. The TV ranged from 0.1 to 13.2 $cm^3$ with a mean value of 2.8 $cm^3$ for a number of indications and prescription doses. The EI (n50%) values were compared to the widely used indices PCI and GI, as changes in these indices would be associated with an increase in the value of EI($n_{50\%}$). In RapidMiner Go the data preprocessing is being automatically processed. The total amount of features required was tested by epochs iteration and it is found that total number of features acquired by DVH that is Volume (cc), Prescribed Dose (Gy), Prescribed Isodose (percent), Coverage, Selectivity, Gradient Index, and PCI were all tested iteratively and were found to be adequate and efficient to predict the EI value. Taking account of all the features the data was split into two sets, one set of 60 plans for training the regression models, as this set contains known output results so the model takes the results to learn in order to generalized it on another set later on for predicting. Another set of 40 plans to test and validate.

It is known from previous study that supervised regression machine learning algorithms are suitable for selecting appropriate models to represent the prediction since we have calculated results [12]. The algorithms used in this analysis were linear regression [13], tree regression [14], SVM [15, 16], GTB, DL, DT, and RFT. They were tested to see how well they could predict the EI value. To evaluate the performance of models we compared its result of $R^2$, Average absolute error and Average relative error, squared correlation error and finally with its robustness of modelling time taken to predict by RapidMiner Go is shown in table 2:

## Results:   Model Performance Metrics:

| Model | Root mean square... | Average absolute e... | Average relative er... | Squared correlatio... | Model building time |
|---|---|---|---|---|---|
| Generalized Linear Model | 0.01 | 0.008 | 1.74 | 0.974 | 2.812 s |
| Deep Learning | 0.011 | 0.009 | 1.81 | 0.979 | 1.663 s |
| Decision Tree | 0.021 | 0.018 | 3.9 | 0.812 | 0.483 s |
| Random Forest | 0.021 | 0.018 | 3.79 | 0.769 | 1.296 s |
| Gradient Boosted Trees | 0.018 | 0.015 | 3.12 | 0.872 | 15.894 s |
| Support Vector Machine | 0.012 | 0.011 | 2.25 | 0.94 | 11.499 s |

**Table 1**. Performance of Regression machine learning algorithm models obtained from RapidMiner Go (Java Programming). Table summarizing the Root mean square error, squared correlation, Average absolute and relative error, predicting speed of modelling time of models created. It used seven dose-volume based parameters from each DVH.

## Comparison of predictions with actual data.

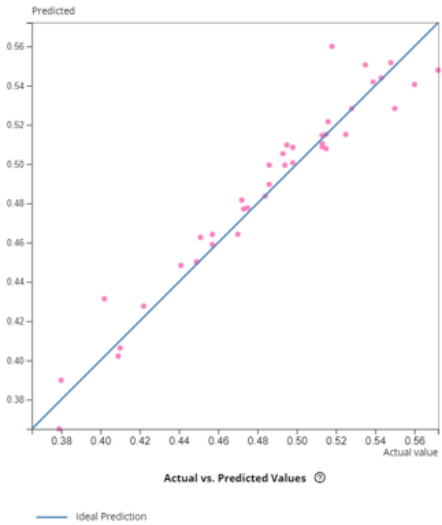
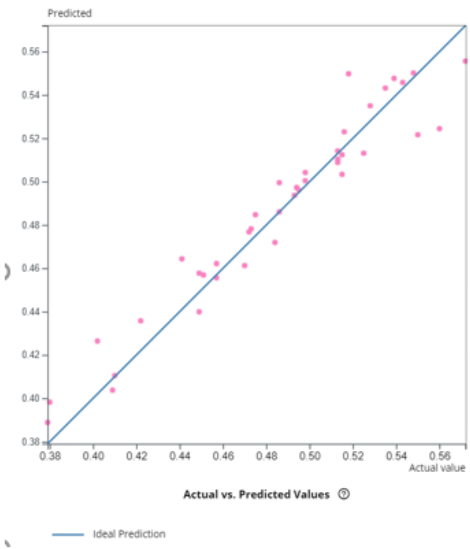
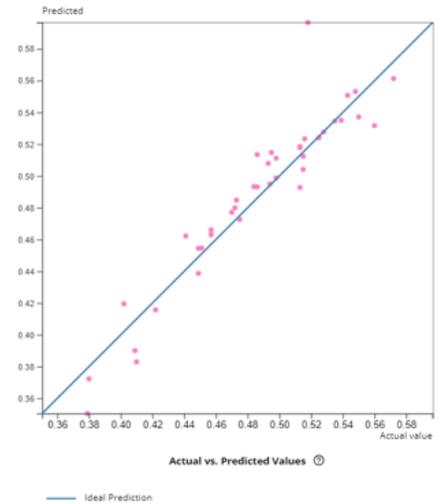

**Generalized Liner Model**  **Deep learning Model**  **Support Vector Machine Model**

*Figure 2. Plot of Actual values versus predicted values plot of EI($n_{50\%}$) values using the (a) GLM (b) DLM (c) SVM regression models by RapidMiner GO machine learning algorithms and statistical tool box.*

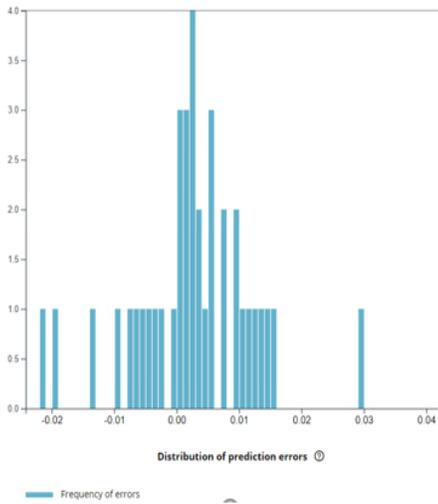
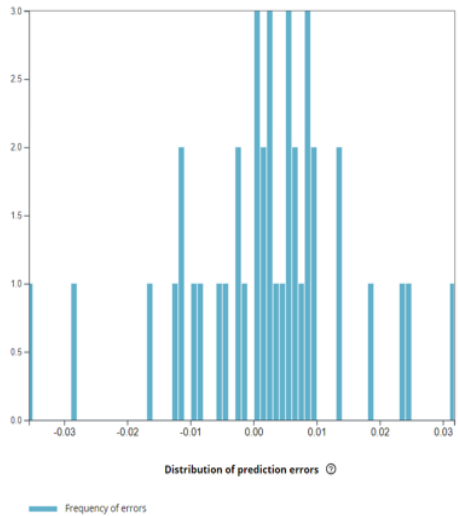
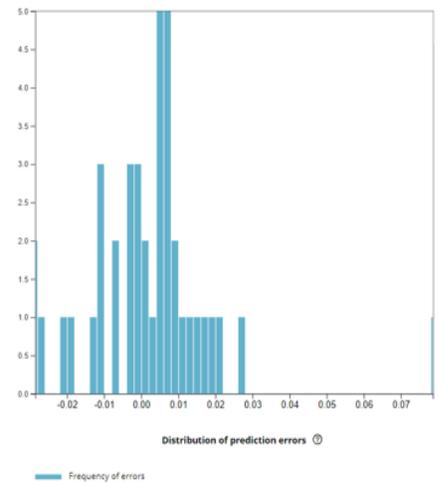

**Generalized Liner Model**  **Deep learning Model**  **Support Vector Machine Model**

*Figure 3. frequency of errors displayed as distribution of predicted errors plot of EI values using the (a) GLM (c) DLM (d) SVM regression models by RapidMiner GO machine learning algorithms and statistical tool box.*

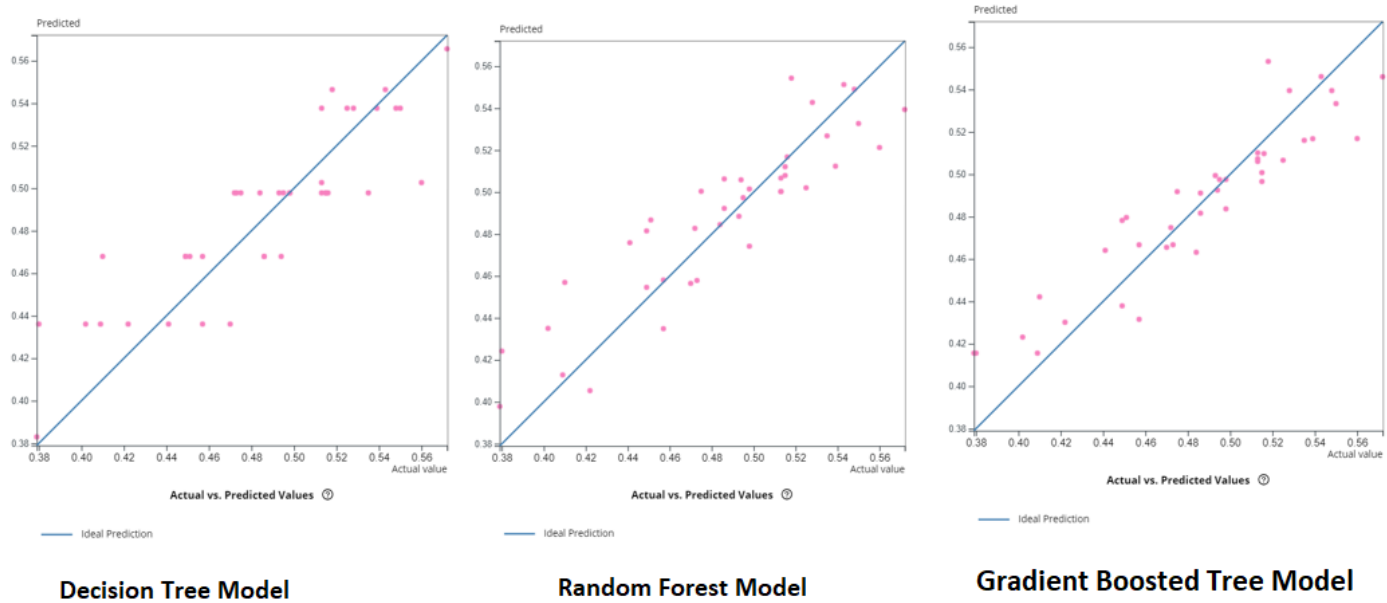

Figure 4. Plot of Actual values versus predicted values EI($n_{50\%}$) using the (a) DTM (b) RFM (c) GBTM regression models by RapidMiner GO machine learning algorithms and statistical tool box.

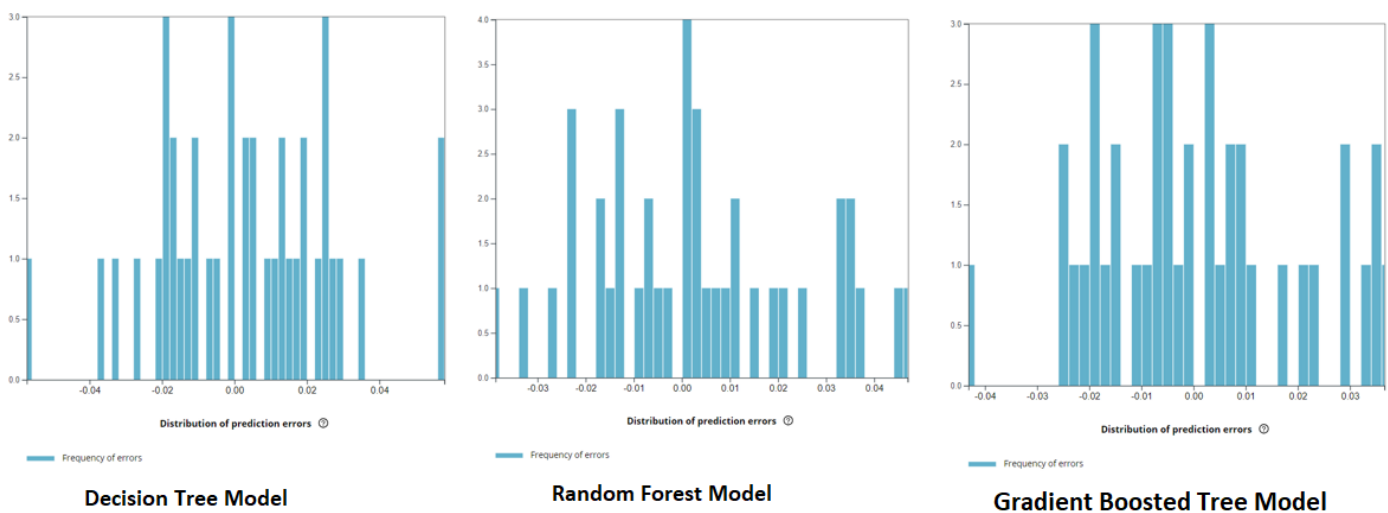

Figure 5. frequency of errors displayed as distribution of predicted errors plot of EI values using the (a) DTM (c) RFM (d)GBT regression models by RapidMiner GO machine learning algorithms and statistical tool box.

## Discussion

the $R^2$ statistics measures how well the model predictions approximate the real data points. An $R^2$ of 1 indicates that the regression predictions perfectly fit the data. GLM accomplished an average absolute error of 0.008. This means if you predict EI:0.48 as a value, the real value is likely between 0.472 and 0.488. $R^2$ statistic is 0.974. SVM model accomplished an average absolute error of 0.011. This means if you predict EI:0.48 as a value, the real value is likely between 0.469 and 0.491. $R^2$ statistic is 0.94. Generally, this is seen as a good value. DL model accomplished an average absolute error of 0.009. This means if you predict EI:0.48 as a value, the real value is likely between 0.4721and 0.489. $R^2$ statistic is 0.979.

Decision Tree model accomplished an average absolute error of 0.018. This means if you predict EI:0.48 as a value, the real value is likely between 0.46199997and 0.498. $R^2$ statistic is 0.812. Random forest accomplished an average absolute error of 0.018. This means if you predict EI:0.48 as a value, the real value is likely between 0.461999997 and 0.498. $R^2$

statistic is 0.769. GBT model accomplished an average absolute error of 0.015. This means if you predict EI:0.48 as a value, the real value is likely between 0.46499997 and 0.495. $R^2$ statistic is 0.872.

From above results of graphical representations and errors table, it is found that the GL regression model produced the best predictive results among all other models, having RMSE error of 0.01 with the modelling time (MT) 2.812 seconds, followed with the DL and SVM producing second best results of predictive models with RMSE 0.11 and 0.012. Modelling time for DL is 1.66 seconds lesser than GL but SVM had the second highest MT among all the models with 11.499 seconds, SVM model were accurate results with compare to GL and with lesser modelling time when it had lesser values of DVH parameters to predict, this means it couldn't perform well with a greater number of parameters. The importance of modelling time increases when the higher number of treatment plans to compare. The DT,RF and GBT were accurate enough to predict since RMSE is found to be 0.021, 0.021 and 0.18, though the MT in DT is lesser than all models but it turn out to be absolute unacceptable for predicting EI.

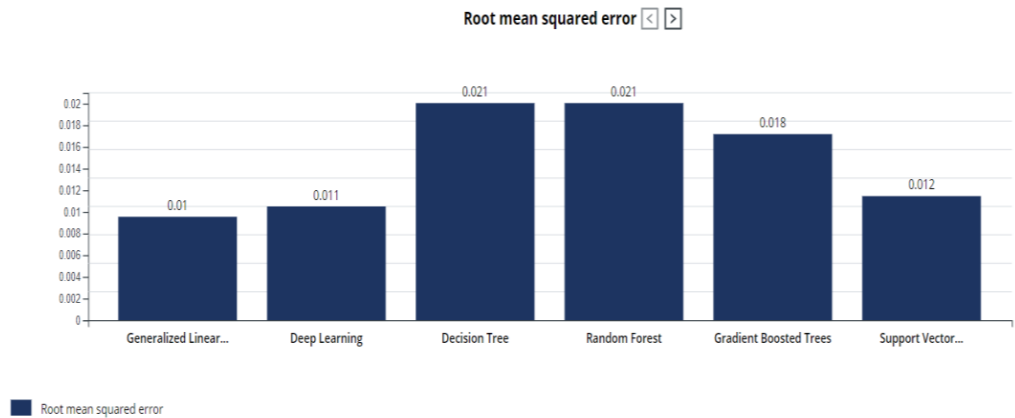

*Figure 6: comparison table of $R^2$ of models used.*

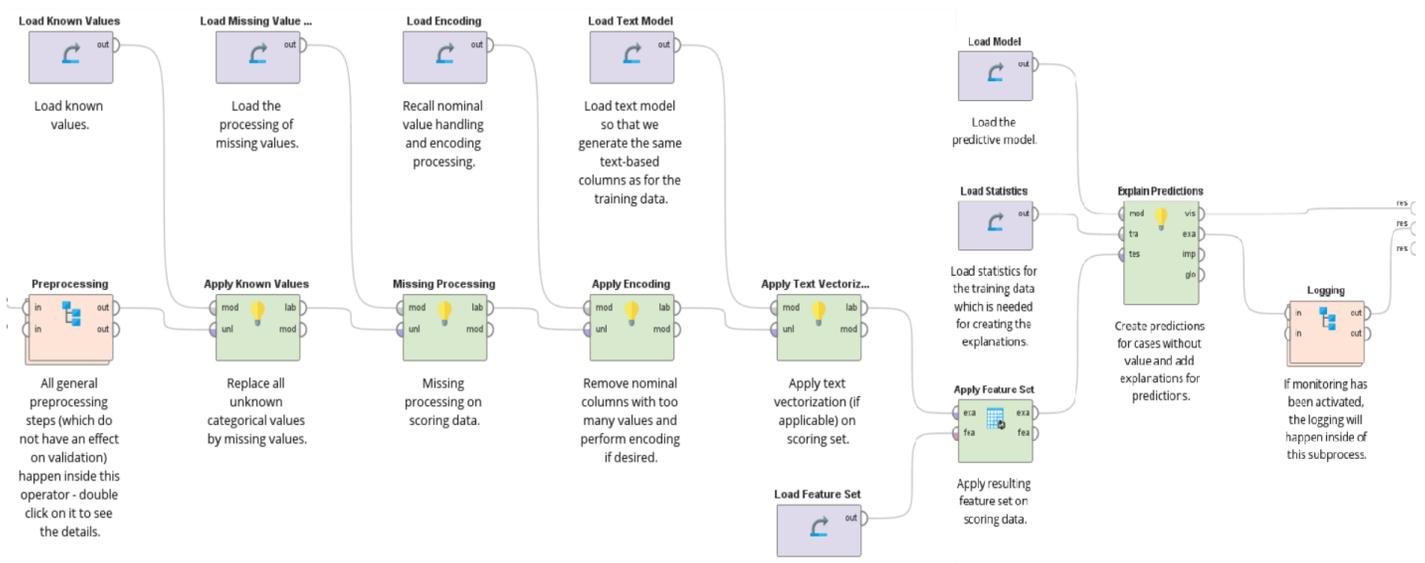

***Figure 6: Flow chart for Validation process of Generalized linear Model in RapidMiner GO (9.8 version), as it explains the connectivity to the attributes and execution of the model to predict.***

*Model Performance Metrics:*

|  | GLM | DL | DT | RF | GBT | SVM |
|---|---|---|---|---|---|---|
| Parameters |  |  |  |  |  |  |
| Selectivity | 0.65 | 0.46 | 0.31 | 0.44 | 0.49 | 0.16 |
| PCI | -0.01 | 0.18 | -0.01 | 0.31 | 0.1 | 0.39 |
| Coverage | -0.06 | -0.14 | 0.01 | -0.01 | 0.01 | -0.15 |
| PI (%) | -0.2 | -0.19 | 0.02 | -0.09 | -0.13 | -0.17 |
| Volume(cc) | -0.07 | -0.07 | -0.04 | 0.1 | -0.03 | -0.13 |
| PD (Gy) | -0.01 | -0.01 | -0.05 | -0.08 | -0.04 | -0.01 |
| GI | -0.76 | -0.85 | -0.75 | -0.68 | -0.79 | -0.86 |

Table 2: IMPORTANT FACTORS FOR PREDICTION: ACCORDING TO MODELS

*The above table shows the most important parametre used is slecetivity for all the models, where as GLM had 0.65 which is the highest among the models. Selectivity is major factor as it defined by treated target volume overthe Prescription isodose Volume. The table 3 explains about the weights of the parameter columns used by all the models for the predictions. GI and selectivity turned out to be the most weighted factors for prediction followed with PCI.*

| Column | Weight |
|---|---|
| GI | 0.799 |
| Selectivity | 0.799 |
| PCI | 0.788 |
| PI(%) | 0.397 |
| PD(Gy) | 0.388 |
| Volume (cc) | 0.272 |
| Coverage | 0.13 |

Table 3. Data Metrics; Weights by correlation of all models analysed: Importance based on correlation with target column.

# Conclusions

*Different Models generated by Machine Learning Algorithms of RapidMiner Go based automated predictive models were used to predict the EI from DVH-extracted parameters of SRS treatment plans. Generalized linear model (GLR), Decision Tree Model, Support Vector Machine model (SVM), Gradient Boosted Trees model (GBT), Random Forest model (RF) and Deep learning Model (DL) models were analyzed. It is found that all models were found to be accurate and very low in errors predictions of EI, except the Random forest and Decision Tree model. So, it is possible to predict EI using RapidMiner Go. It would be worth to compare the algorithms with Python and MATLAB programming using the same models with the manual technique such as Supervised learning.*